\documentclass[conference]{IEEEtran}
\ifCLASSINFOpdf
\else
\fi
\usepackage{amsmath}
\usepackage{graphicx}
\usepackage{algorithm}
\usepackage{algorithmic}
\usepackage{multicol}
\usepackage{lipsum}
\usepackage{subcaption}

\makeatletter
\newcommand\fs@norules{\def\@fs@cfont{\bfseries}\let\@fs@capt\floatc@ruled
  \def\@fs@pre{}%
  \def\@fs@post{}%
  \def\@fs@mid{\kern3pt}%
  \let\@fs@iftopcapt\iftrue}
\makeatother
\floatstyle{norules}
\restylefloat{algorithm}

\begin{document}

\title{Network Analysis as a Tool for Shaping Conservation and Development Policy: A Case Study of Timber Market Optimization in India}

\author{
    \IEEEauthorblockN{
        Xiou Ge\IEEEauthorrefmark{3}\textsuperscript{\textsection}, Sarah E. Brown\IEEEauthorrefmark{2}\textsuperscript{\textsection}, 									Pushpendra Rana\IEEEauthorrefmark{2}, 
        Daniel C. Miller\IEEEauthorrefmark{2},
        Lav R. Varshney\IEEEauthorrefmark{3}}
    \IEEEauthorblockA{\IEEEauthorrefmark{3}Department of Electrical and Computer Engineering}
    \IEEEauthorblockA{\IEEEauthorrefmark{2}Department of Natural Resources and Environmental Sciences}
    \IEEEauthorblockA{University of Illinois at Urbana-Champaign
    \\\ Urbana, Illinois, 61801}
    \IEEEauthorblockA{\{xiouge2, saraheb3, prana5, dcmiller, varshney\}@illinois.edu}
}

% make the title area
\maketitle

\begingroup\renewcommand\thefootnote{\textsection}
\footnotetext{Equal contribution}
\endgroup

% As a general rule, do not put math, special symbols or citations
% in the abstract
\begin{abstract}
The incorporation of trees on farms can help to improve livelihoods and build resilience among small-holder farmers in developing countries. On-farm trees can help generate additional income from commercial tree harvest as well as contribute significant environmental benefits and ecosystem services to increase resiliency. Long-term benefits from tree-based livelihoods, however, depend on sustainable timber harvesting. In this paper, we discuss the potential for network analysis as a tool to inform conservation and development decision-making. Specifically, we formulate the commercial tree market between farmers and traders as a transportation problem and optimize the transactions. We create a model of the commercial tree market in the Bilaspur district of Himachal Pradesh, India based on a detailed dataset of market interactions between farmers and timber traders, using the existing road network of this region. Using this model, we perform a maximum-flow-minimum-cost optimization for tree commodity flow. We compare the results of our optimized model with actual flow within the network, and we find a high potential to increase efficiency of market transactions within this region, noting a significant reduction to the minimum-cost flow value for our optimized model compared to the flow cost for actual transactions. 
We propose that using this network flow optimization model to strategically distribute permits can reduce costs associated with market transactions. Our results suggest that this direct policy action would be beneficial to the region. Finally, we suggest that cost savings could be used to establish tree planting programs to support a long-term sustainable tree market. Shaping policies to address these market inefficiencies in developing regions could help support and elevate tree-based livelihoods for farmers, traders, and industries.\end{abstract}

\section{Introduction}

\IEEEPARstart{T}{imber} harvest from on-farm trees represents a significant path to mitigate poverty and improve livelihoods among small-holder farmers in developing regions, such as the Indian Himalayas \cite{Kalaba2010,Miller201747}. Trees-on-farms contribute significant environmental benefits and ecosystem services, such as soil stabilization, runoff prevention, hydrological cycling, and biodiversity conservation, along with the socio-economic benefits realized through additional income sources \cite{Miller201747,REED201762}. However, tree-based livelihoods depend upon sustainable timber harvesting for long-term benefits to occur. 

We consider the tree market transactions between farmers and traders in the Bilaspur District, Himachal Pradesh, India. This region has a program of planned rotational harvesting to supply markets with khair (\textit{Acacia catechu}) trees while allowing natural regeneration of farmer's trees to occur between harvest cycles. The Himachal Pradesh Forest Department (HPFD) regulates the harvesting of khair trees on smallholders’ woodlots. This regulation is governed by the Land Preservation Act of 1978 (LPA 1978), which requires a 10-year felling cycle to allow tree regeneration and limit soil erosion \cite{LPA1978}. The region is divided into forest blocks, and each block is eligible to be harvested only once every ten years. Governments issue harvesting permits to traders based on the number of trees marked on farmer's land, which limits how much each trader can harvest.
Farmers in this region grow small patches of commercial khair trees on their farms, and traders come to the farms to harvest the trees to sell to markets. The industries supplied by these markets use the exudate from the heartwood of khair trees to make catechu, which is the main ingredient in several medicines, tannins, dyes, and mouth fresheners. These products have a well-established market in India and are sold commercially throughout India. 

However, the sustainability of this market system is currently threatened due to a high prevalence of destructive tree harvest. Traders have taken to uprooting trees during harvest to increase their cash flow and meet market demands. However, uprooting trees is a destructive procedure, which prohibits the natural regeneration cycle of trees within this rotational system. With uprooting harvest techniques, farmers forgo future potential revenue from tree regrowth for increased immediate payments. In doing so, they also relinquish the ecological benefits of trees on farms, increasing soil erosion and land degradation. As in many developing countries, we see that even with governmental regulations in place, there is no guarantee that local communities will benefit from the use of their natural resources or maintain a sustainable harvest cycle \cite{Pacheco2012}. 

In this paper, we use a detailed dataset of the farmer to trader transactions occurring from 2009--2014 in Bilaspur District. We model the flow of timber between farmers and traders in this region, based on the coordinate locations of each farmer and each trader and the distances along a detailed road network between each farmer and each trader. We suggest network analysis as a valuable tool for capturing market transaction inefficiencies, such as those in transportation costs. The maximum-flow-minimum-cost problem in network analysis offers a technique to model flow through a network and is commonly used as an optimization technique for supply-demand flow problems \cite{Ahuja:1993:NFT:137406}. We use this method as an approach to capture the inefficiencies in the tree harvest flow network and suggest pathways to promote sustainable tree harvest and local livelihoods. We compare our optimization model to the actual data to show the cost reduction potential that could provide a pathway to more sustainable tree management. We conclude that the current distribution of permits leads to unnecessarily high transportation costs, and we calculate an optimal distribution of permits based on minimizing the cost of timber flow through the market network. 

We believe this model can inform direct policy action in this region. We suggest our model as a tool to strategically plan permit distribution among the traders and establish village-level tree planting programs. A more efficient approach to distributing permits can help prevent delays to harvesting trees within the cycle schedule, which is a driver of uprooting due to reduced monitoring of harvesting occurring beyond the normal harvest cycle for an area. Furthermore, there is the potential for significant cost savings associated with transportation by implementing this optimized model. These cost savings could be used to benefit local communities and rural development. We suggest the promotion of village-level tree planting projects, and model which villages should be the highest-priority targets for such a program, based on our algorithm to achieve maximum-flow at minimum-cost while maintaining an equitable distribution of trees in all villages.

% You must have at least 2 lines in the paragraph with the drop letter
% (should never be an issue)

%%\hfill 
 
%%\hfill May 2, 2018
\section{Dataset}
Our dataset covers the period from 2009--2014 in the Bilaspur District of the Himachal Pradesh state in the Himalayas, India. The data consists of information on 9,481 market transactions among 5,911 individual farmers from 304 villages and 154 timber traders in the region, and it captures all legal market transactions during this time period. The dataset documents the permits allocated to traders by the state as well as the number and volume of trees actually harvested by the trader, including number of uprooted trees harvested. Each interaction within the dataset has farmer-level data on farm size, trees harvested and sold, socio-economic variables as well as information on the trader, the village, the forest beat, and the market to which the trader is selling the tree product. Over 200 variables for each transaction are recorded.

We model transaction costs between villages and traders as the distance between the village and the trader along the road networks in this region. The road network data pertains to the same time period as our dataset; however, we note that there are a small number of traders coming into Bilaspur district from other districts. The detailed road network does not extend beyond Bilaspur district, so we use lower resolution global roadmap data to connect these distant traders into the Bilaspur road network. 

\begin{figure}
\centering
\includegraphics[width=1\columnwidth]{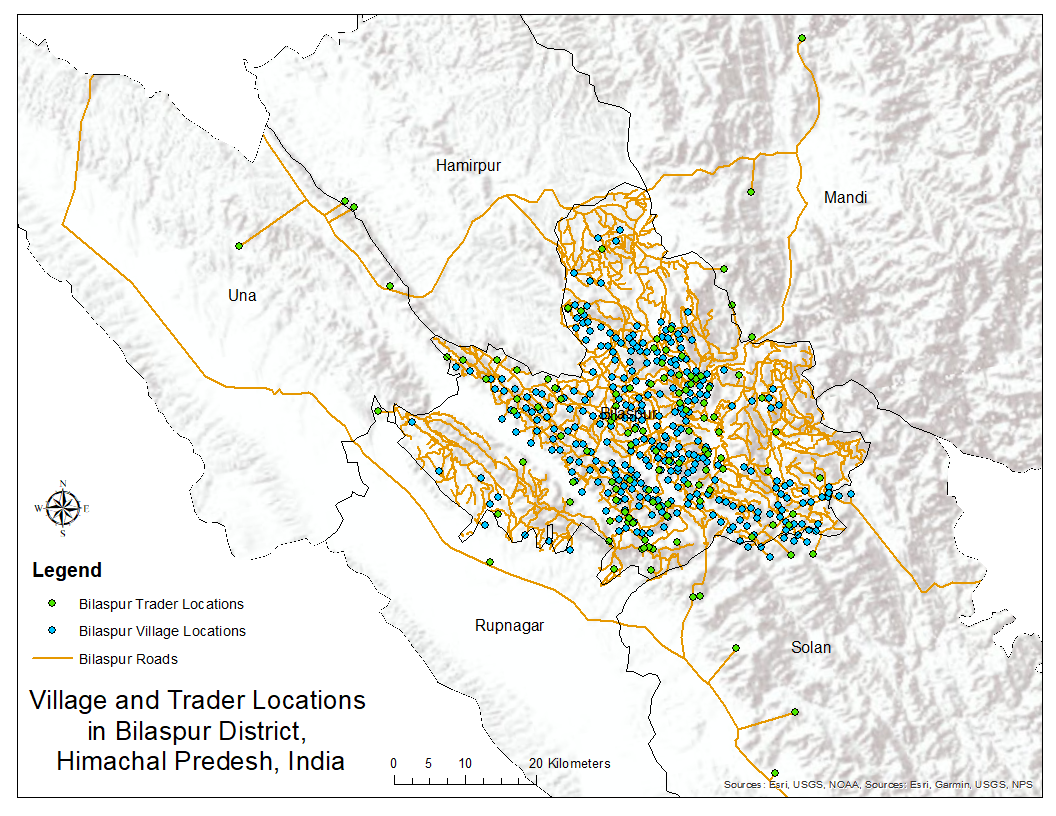}
\caption{Map of Bilaspur District, Himachal Pradesh, India.}
\label{fig:Map}
\end{figure}

For our model, we aggregate the farmer-level transactions onto the village-level. We do this because village-level results are more interpretable and more relevant for determining policy implications. Farmer-level analysis would require more information on farmer decision-making and social ties to offer useful results. We argue that village-level analysis is a good first step since there can be clear policy actions for interventions at the village-scale. We acknowledge that farmer-level analysis is important for understanding market equality and marginalization of individual farmers within-villages. We argue, however, that promoting sustainable harvest strategies at the village-level will help improve livelihoods at the level of the entire population. Fig.~\ref{fig:Map} shows the map of villages, traders, and the road network for Bilaspur District.

\section{Optimization Model}
\subsection{Theory}
\begin{figure}[h!]
\centering
\includegraphics[width=1\columnwidth]{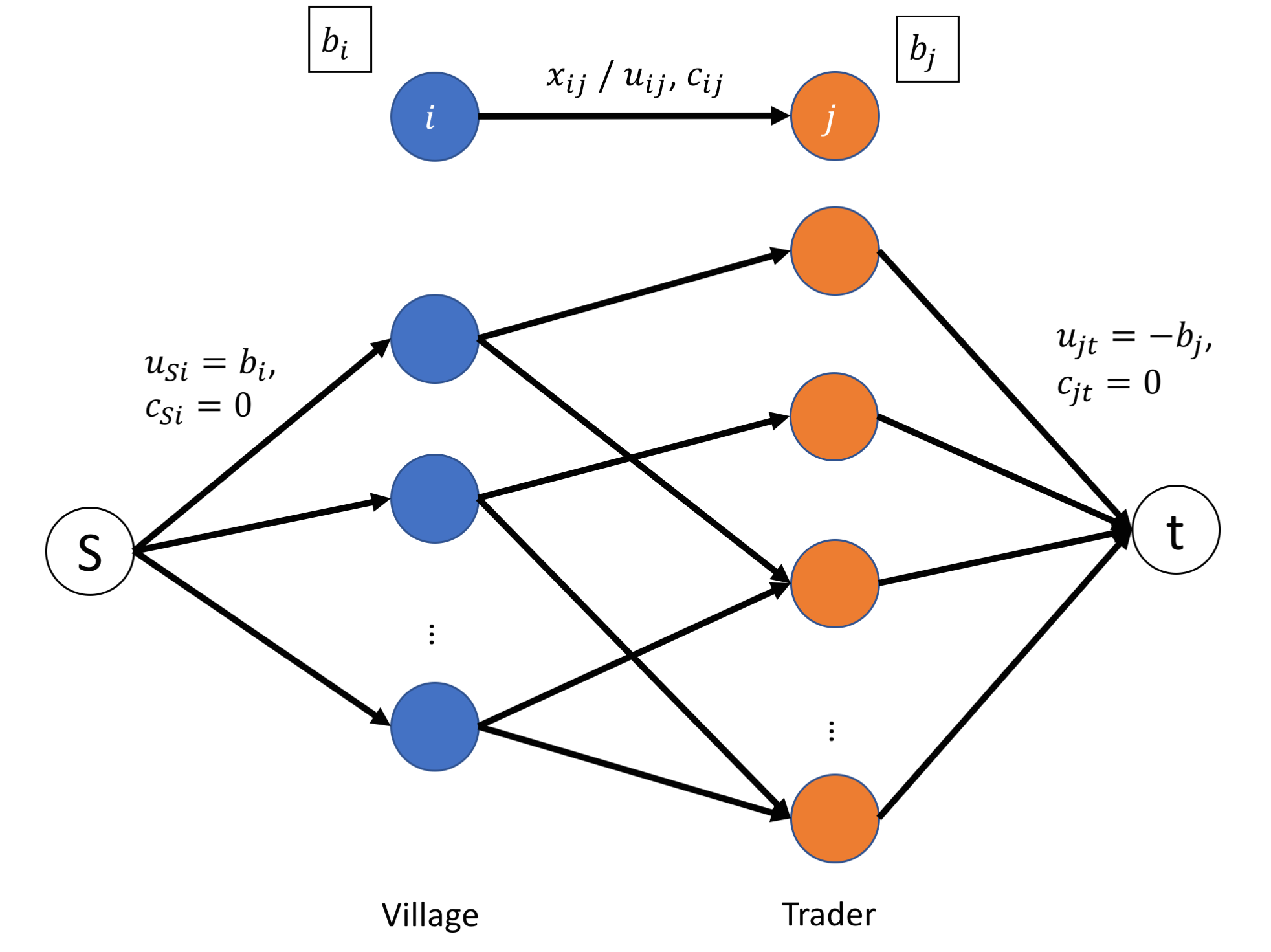}
\caption{Transformed max-flow-min-cost network model.}
\label{fig:old_connections}
\end{figure}

We take a similar flow-based modeling approach to our problem as the racial balancing of schools problem presented by Ahuja et al. \cite{Ahuja:1993:NFT:137406}. There are potentially three groups of nodes: farmers, villages, and traders. Farmers in each village supply trees, whereas traders demand trees. Ideally, the supply should meet the demand without uprooting any trees. The tree harvest transactions described above can be modeled as flows from supply nodes to demand nodes with flow costs as the distance between a supply node (village) and demand node (trader) along the road network required to transport a unit of harvested trees, as shown in Fig.~\ref{fig:old_connections}.

Let $G=(V,E)$ be a directed network defined by a set $V$ of nodes and $E$ of directed edges $(i,j)\in A$. Each edge $(i,j)\in A$ has an associated cost per unit flow $c_{ij}$, flow variable $x_{ij}$, and a flow capacity $u_{ij}$. We assume the lower bound on flow for all edges to be 0 for our problem. We use $b(i)$ as an integer number to represent the supply/demand associated with node $i$. If $b(i)>0$, node $i$ is a supply node; if $b(i)<0$, node $i$ is a demand node with a demand of $-b(i)$. The objective function for the optimization problem can be formulated as follows:
\begin{equation}
\begin{aligned}
& {\text{minimize}}
& & \sum_{(i,j)\in E} c_{ij}x_{ij} \\
& \text{subject to}
& & \sum_{j:(i,j)\in E} x_{ij} - \sum_{j:(j,i)\in E} x_{ji} = b(i) & \forall i \in V, \\
&&& 0\leq x_{ij} \leq u_{ij} & \forall (i,j) \in E.
\end{aligned}
\end{equation}

There are many approaches to solve the minimum-cost problem. One approach is to first find the max-flow of the transformed network without considering the cost, and then use the negative canceling algorithm to find the cost minimized solution. To obtain the transformed network, we introduce a source node $s$ and a sink node $t$. For each node $i\in V$ with $b_i > 0$, add a source edge $(s,i)$ to $G$ with capacity $b_i$ and cost $0$. For each node $i\in V$ with $b_i < 0$, add a source edge $(i,t)$ to $G$ with capacity $-b_i$ and cost $0$. 

The max-flow objective is stated as follows:
\begin{equation}
\begin{aligned}
& \text{maximize} & & v \\ 
& \text{subject to}
& & \sum_{j:(i,j)\in E} x_{ij} - \sum_{j:(j,i)\in E} x_{ji} \\ & & & = 
              \left\{
                \begin{array}{ll}
                  v &\text{for} \;i=s\\
                  0 &\forall i \in V -\{ s \;\text{and} \;t\}\\
                  -v &\text{for} \;i=t
                \end{array}
              \right.
 \\
&&& 0\leq x_{ij} \leq u_{ij}, \; \forall (i,j) \in E.
\end{aligned}
\end{equation}

To solve the max flow problem, we apply the Ford-Fulkerson algorithm \cite{FordF1956}. Then, after running the Ford-Fulkerson algorithm, we obtain the flow network and residual network. By taking cost into consideration, we find a directed cycle in the residual network that has negative net cost. We apply the cycle-canceling algorithm to reduce the total cost of a feasible flow, as in \cite{Ahuja:1993:NFT:137406}.

\subsection{Village-Trader Model of Max-Flow-Min-Cost}
We model the 304 villages, which consist of 5,911 farmers, and the 154 traders within the region. We consider each village as a supply node and approximate the supply capacity by adding up the volume of trees harvested from all plots within the village. To aggregate our farmer-level data onto villages, we sum the tree capacity of farmer's land within each village. For the data-based model, this is simply the sum of the trees sold within the time period by all the farmers in that village. For our optimized model, we assume that each farm has a carrying capacity based on the average yield for the land use type of the farm and the area of the farm. We do this to capture the potential supply of each village, under the assumption that certain villages have been over-harvested or under-planted, and there is a potential to have a more equitable landscape. 

\begin{figure}[h!]
  \centering
  \includegraphics[width=0.5\textwidth]{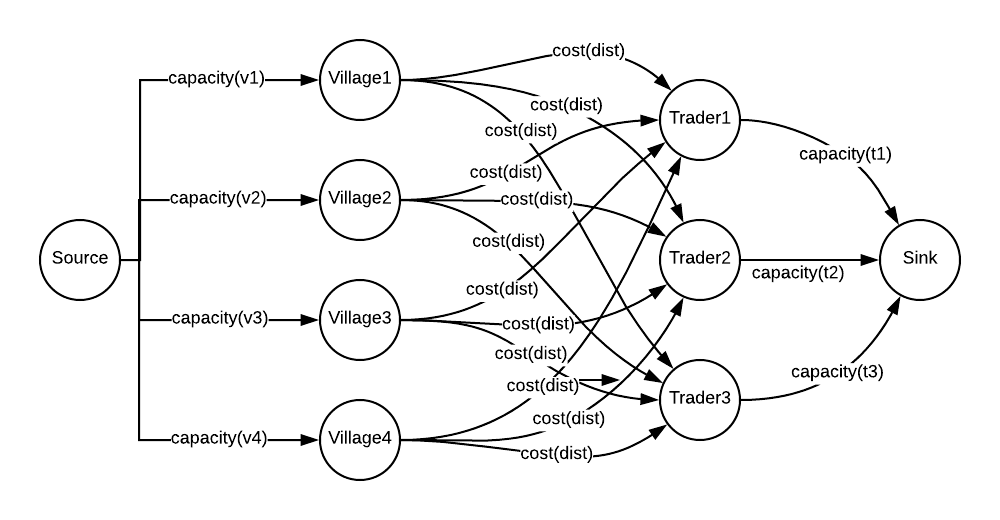}
  \caption{Village-Trader max-flow-min-cost model diagram.}
  \label{fig:FlowModel}
\end{figure}

We calculated the average yield for each land use type from the data by separating farmers by each land use type and averaging for each type the observed number of trees harvested from the farm divided by the production area of the farm. From this, we get an estimate of the per hectare mean yield for each land use type. Let $Y$ be a vector of the yield of that particular land type, $T$ be a vector of the land type of each farm, $S$ be a vector of area of each farm, and $M$ be the set of farms locate in a particular village. Then the tree supply from a particular village node can be computed as: 

\begin{equation}
b_v=\sum_{i\in M}{Y_{T_i}S_i}. 
\end{equation}

The demand of a particular trader node is the sum of trees handled by that trader. Let $H$ be the vector of number of harvested trees in each transaction and $R$ be the set of transactions relevant to a trader.
\begin{equation}
b_t=\sum_{i \in R}H_i
\end{equation}

We set the cost of on the edge between villages and traders nodes as the distance along the road network between each village and trader. The process for generating these values is described in the the following subsection.

We construct this model as a source-sink flow model, shown in Fig.~\ref{fig:FlowModel}, with edge capacities and costs as defined above. The villages and traders are initially fully connected, such that each village is allowed to supply every trader. The maximum-flow-min-cost algorithm is then run for this model. This algorithm returns the flow pathways to maximize flow and minimize cost. The resulting flow returns zeros on many of the edges between villages and traders, indicating that zero flow occurs between them in the optimized model. We remove the zero edges to construct the optimized network shown in Fig.~\ref{fig:new_connections}.

\subsection{Road Network Distance Analysis}
We model transaction costs between villages and traders as the distance between the village and the trader along the road networks in this region. To find these distances, we used ArcMap10.5 to map village and trader locations, shown in Fig.~\ref{fig:Map}. We use the ArcMap Network Analyst extension to employ the OD Cost Matrix analysis tool to calculate the distances via the road network between each point on our village layer to each point on our trader layer and create a table with the village id, trader id, and distance in kilometers between them. The OD Cost Matrix analysis tool generates minimum distances along a network feature between two point layers. We generate a shapefile of villages as the origin point layer and a shapefile of traders as the destination point layer. We transform our road map line layer into a network layer to run the analysis. 

\section{Results}
We find high potential to improve efficiency in market transactions within this region. The optimized model for minimum cost flow resulted in a minimum cost flow value, which is modeled as the total trader-village distance for commodity transfer, of 1,314,671 km. This is compared with the cost flow value returned for actual data of 2,516,444 km, showing a cost reduction of 47 percent. 

\begin{figure}[h!]
\centering
\includegraphics[width=1\columnwidth]{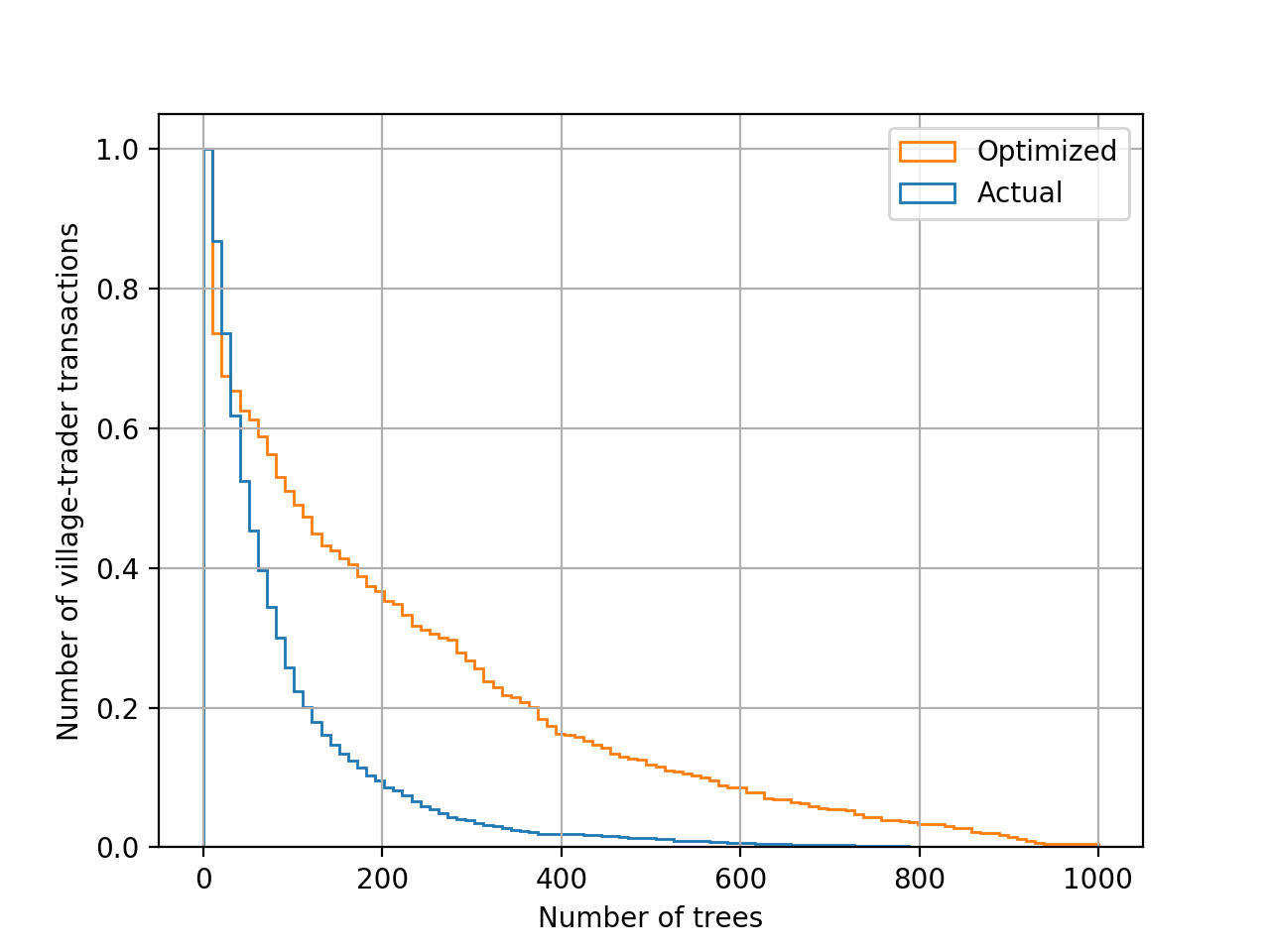}
\caption{Survival function plot comparing the distribution of transactions by their tree harvest volume for the actual transactions and the optimized network flow.}
\label{fig:change_in_transactions}
\end{figure}

In order to achieve this cost savings, commodity flow between villages and traders is optimized, resulting in a more consolidated network where traders harvest from specific villages rather than travel to many dispersed villages to harvest trees. This reduction in dispersed interactions is depicted in Fig.~\ref{fig:change_in_transactions}, showing that many small-volume village-trader transactions are replaced with more large-volume village-trader interactions. 

\begin{figure*}[h!]
\centering
\mbox{
\begin{subfigure}{1\columnwidth}
\includegraphics[width=1\columnwidth]{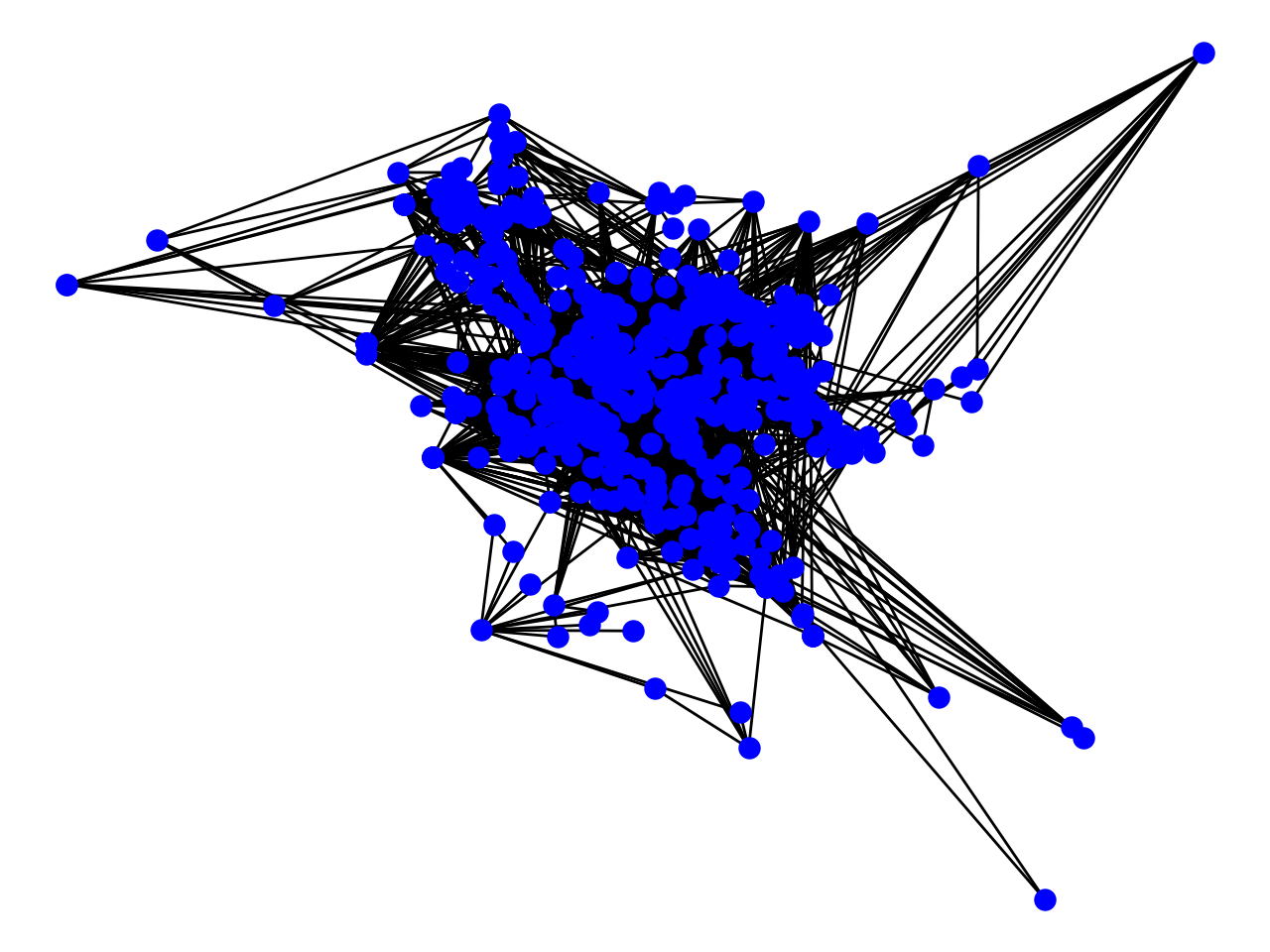}
\caption{Network graph of data set connections}
\label{fig:original_connections}
\end{subfigure}

\begin{subfigure}{1\columnwidth}
\includegraphics[width=1\columnwidth]{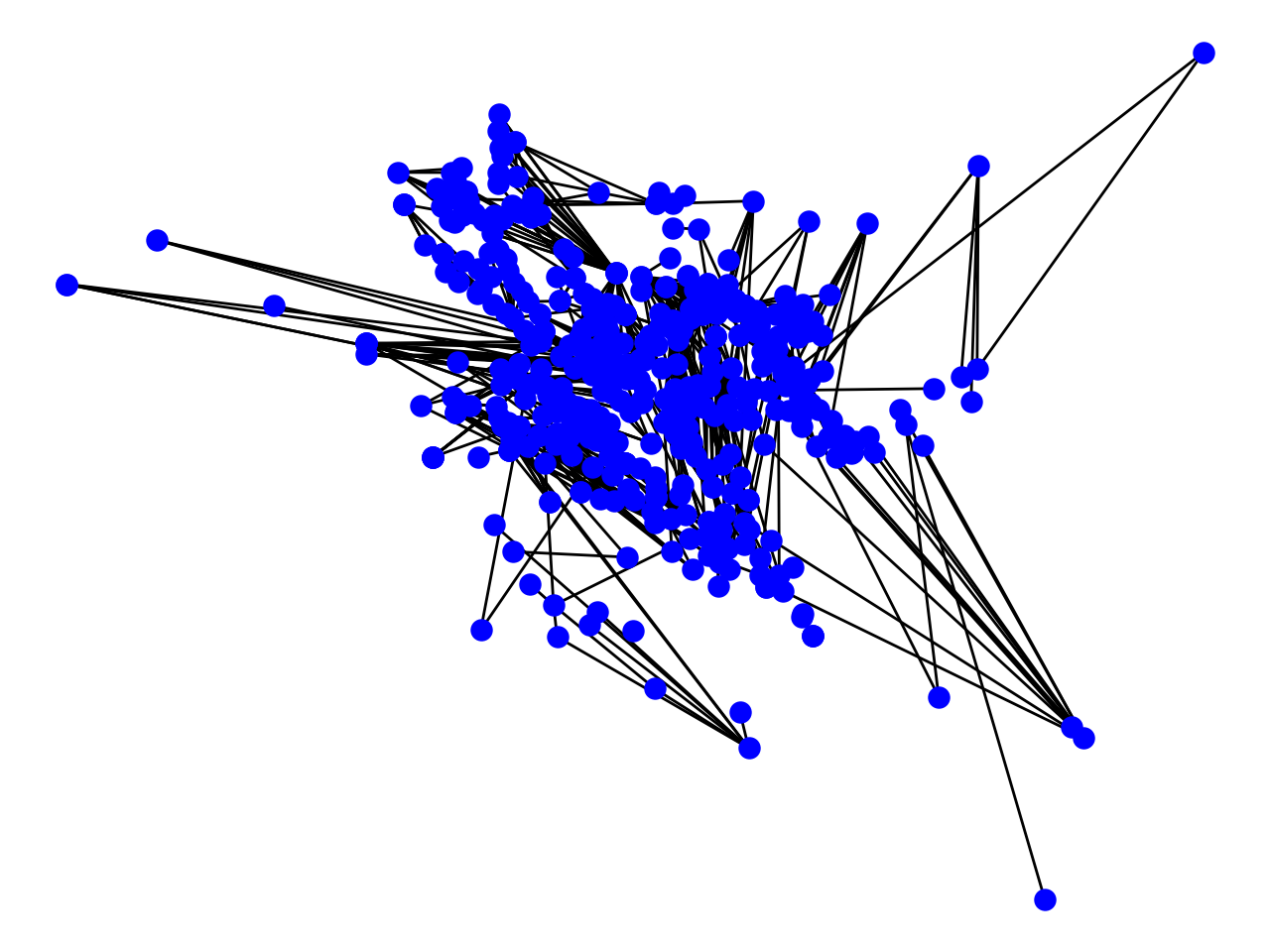}
\caption{Network graph of the optimized network}
\label{fig:new_connections}
\end{subfigure}
}
\caption{Comparison of network graphs between the data and optimized networks.}
\label{fig:connections_comparison}
\end{figure*}

We show network connections from actual transaction data (Fig.~\ref{fig:original_connections}) and the optimized model (Fig.~\ref{fig:new_connections}), which further emphasizes this point. The reduction in edges enables a more efficient flow pathway, which significantly reduces costs associated with transporting harvested trees. 

To achieve better flow, a proposed action may be to strategically distribute permits to traders. The state forest department that designates these permits could use the optimized flow model to assign traders to harvest from specific villages in a more coordinated approach. We further recommend that the cost savings realized from a more efficient harvest scheme could be used to establish tree planting programs to replant previously uprooted trees and sustain levels of trees in landscape to meet market demands. 

\begin{figure*}[h!]
\centering
\mbox{
\begin{subfigure}{1\columnwidth}
\includegraphics[width=1\columnwidth]{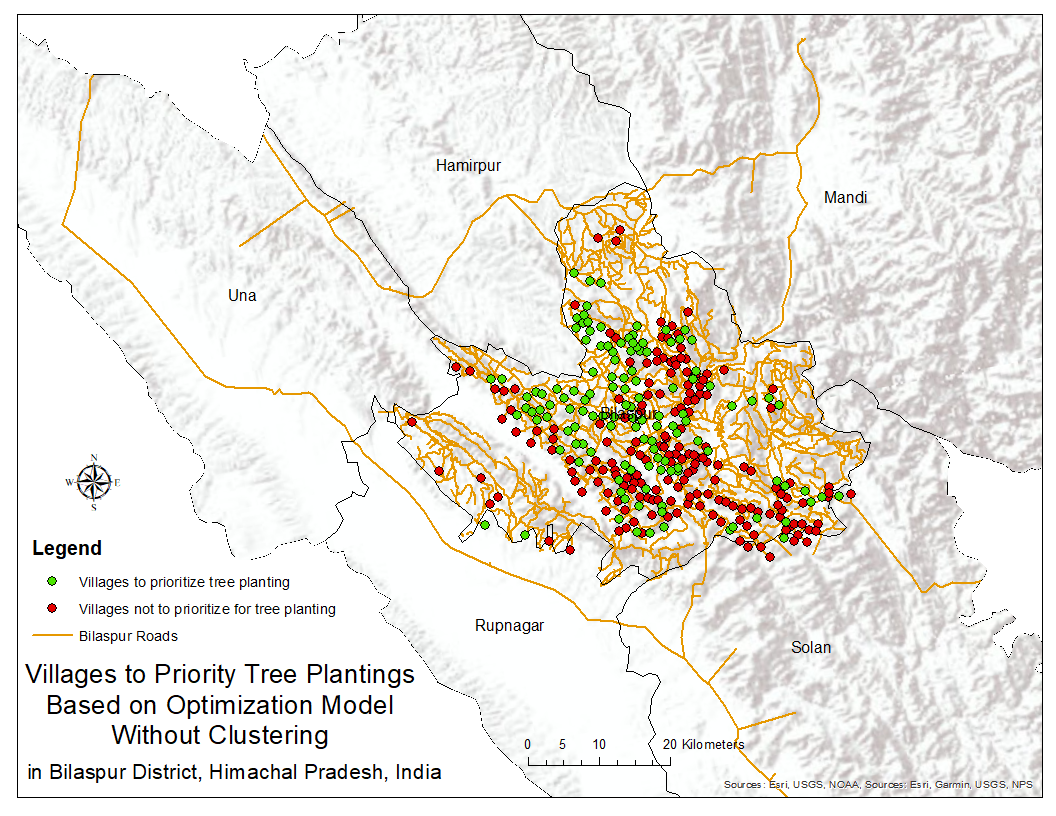}
\caption{Without clustering}
\label{fig:PriorityVillages_without}
\end{subfigure}
\begin{subfigure}{1\columnwidth}
\includegraphics[width=1\columnwidth]{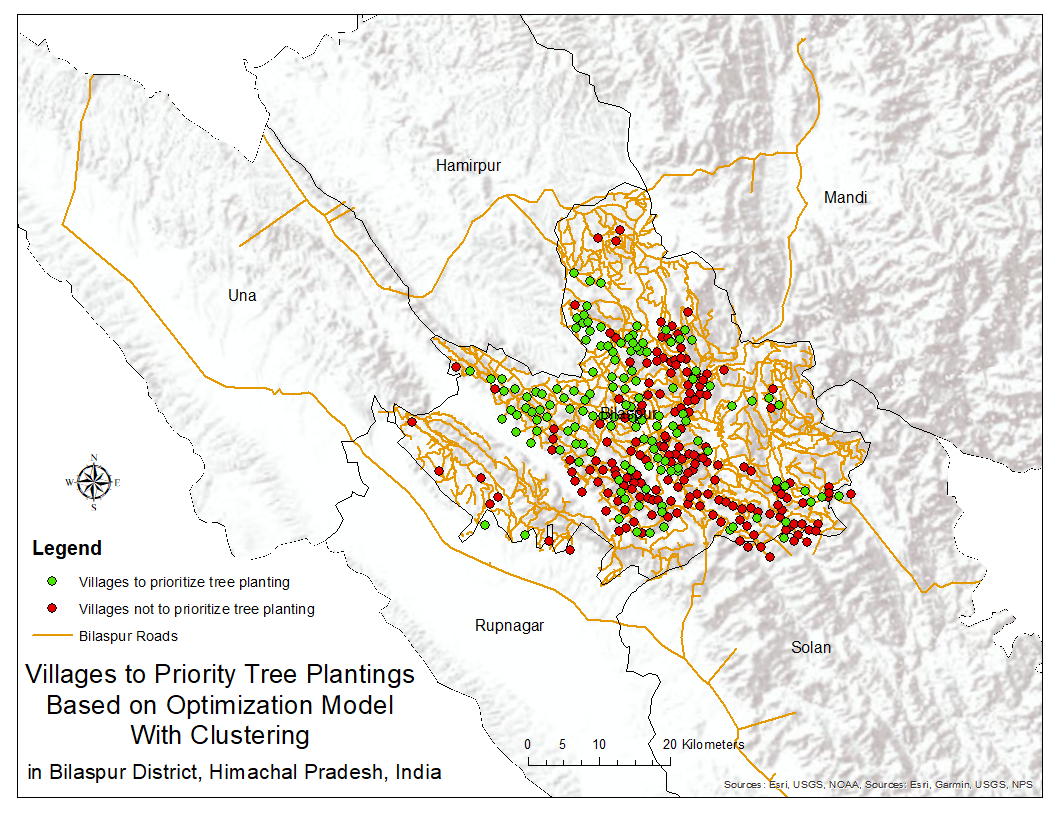}
\caption{With clustering}
\label{fig:PriorityVillages_clustering}
\end{subfigure}
}
\caption{Villages to prioritize for tree planting programs to most efficiently meet market demand based on min cost optimization.}
\label{fig:PriorityVillages}
\end{figure*}

With this recommendation, another result generated by our model is an indication for which villages should be the focus of tree planting programs, determining priority villages where supplies required to minimize cost flow are not currently being met but are determined to be achievable based on the farm area and land use types within the village. We show the high priority (green dots) and low priority (red dots) villages for tree planting/re-planting in Fig.~\ref{fig:PriorityVillages_without}. Promoting the maintenance of a sustainable tree supply to meet market demands through tree planting programs can potentially reduce or eliminate the number of uprooted trees, which was 24,205 during the study period. This solution may help to encourage and maintain farmer and trader tree-based livelihoods. 

\subsection{Clustering}
In order to promote more equitable trade and restrict powerful traders from dominating the market, we design a permit distribution scheme to moderate the demand of traders. Based on transaction history, we classify traders as very low-, low-, medium-, high-, and very high-volume traders based on the total timber volume in all transactions associated with that trader. By applying the hierarchical agglomerative clustering algorithm~\cite{scikit-learn}, we cluster the traders into these 5 groups based on their estimated demand. We then compute the average demand within each cluster and assign permits restricting harvest volume to traders based on their corresponding cluster average. With this restriction imposed, the total transportation cost is further reduced by 42 percent from the initial optimized result to 769,120 km, which reduces the original cost (based on the data) by a factor of 3.27 (or, 70 percent cost reduction). We present the results of our analysis for determining villages that are a priority for tree planting programs in Fig.~\ref{fig:PriorityVillages_clustering}.

\section{Limitations and Future Work}
Although our model offers promising results, there are several limitations and areas for future work. First, we have limited data on the uprooted trees, having only incomplete trader-level data for number of trees uprooted. Future data collection on trees uprooted by farm or by village would help create a better model predicting which villages to target for tree planting programs. Currently, our model assumes that the number of trees needing to be planted for each village is the difference between the calculated optimal number of trees in a village and the actual number of trees harvested in that village. Positive values indicate villages where more trees need to be planted in order to achieve optimal market flow, and negative values indicate villages where current tree production already meets the demands of the market in the optimal flow scenario. A better model would add to this value the observed number trees uprooted in the village, which would address the need for higher levels of tree planting in villages that had more trees uprooted. Since uprooting the trees is destructive, it removes those trees from the landscape, so only trees that were non-destructively harvested can regenerate and be available to meet market demands in future harvest cycles. Since we cannot incorporate the village-level extent of tree uprooting, we assume uniform levels of uprooting across all villages in our current model.

Furthermore, we have limited knowledge on the estimated capacity of the landscape. Our model is based on the mean carrying capacity of the documented land use types. This parameter of the model could be improved by incorporating more informed assumptions and additional variables into the model for land use capacity, such as including elevation in the capacity calculation and variability of farmer socio-economic statuses between villages.

We also suggest developing an improved model of the transport costs of timber between villages and traders. Currently our costs are based on distance between village and trader based on road networks only, but a better measure of accessibility may include quality of roads and terrain. Additionally, it is of interest to incorporate markets into the network model, with actual locations and demands of each market. We have incomplete data on flow to markets, so we were currently unable to include this in our model. A more detailed road map of Bilaspur District as well as an extension of the map to distant traders and markets would enable a more precise analysis.

Finally, it is important to solicit feedback from the policy- and decision-maker perspective on priorities for constraints in the optimization problem. We designed our model to be adaptable and allow users to adjust variables and constraints easily. We acknowledge that ours is only one solution of an optimization problem, and the goals and priorities that define optimization vary between differing perspectives. We offer our current model as a subject for discussion and as a demonstration of the potential suitability of this type of approach to help address sustainability issues in this region.

\begin{figure}
\includegraphics[width=1\columnwidth]{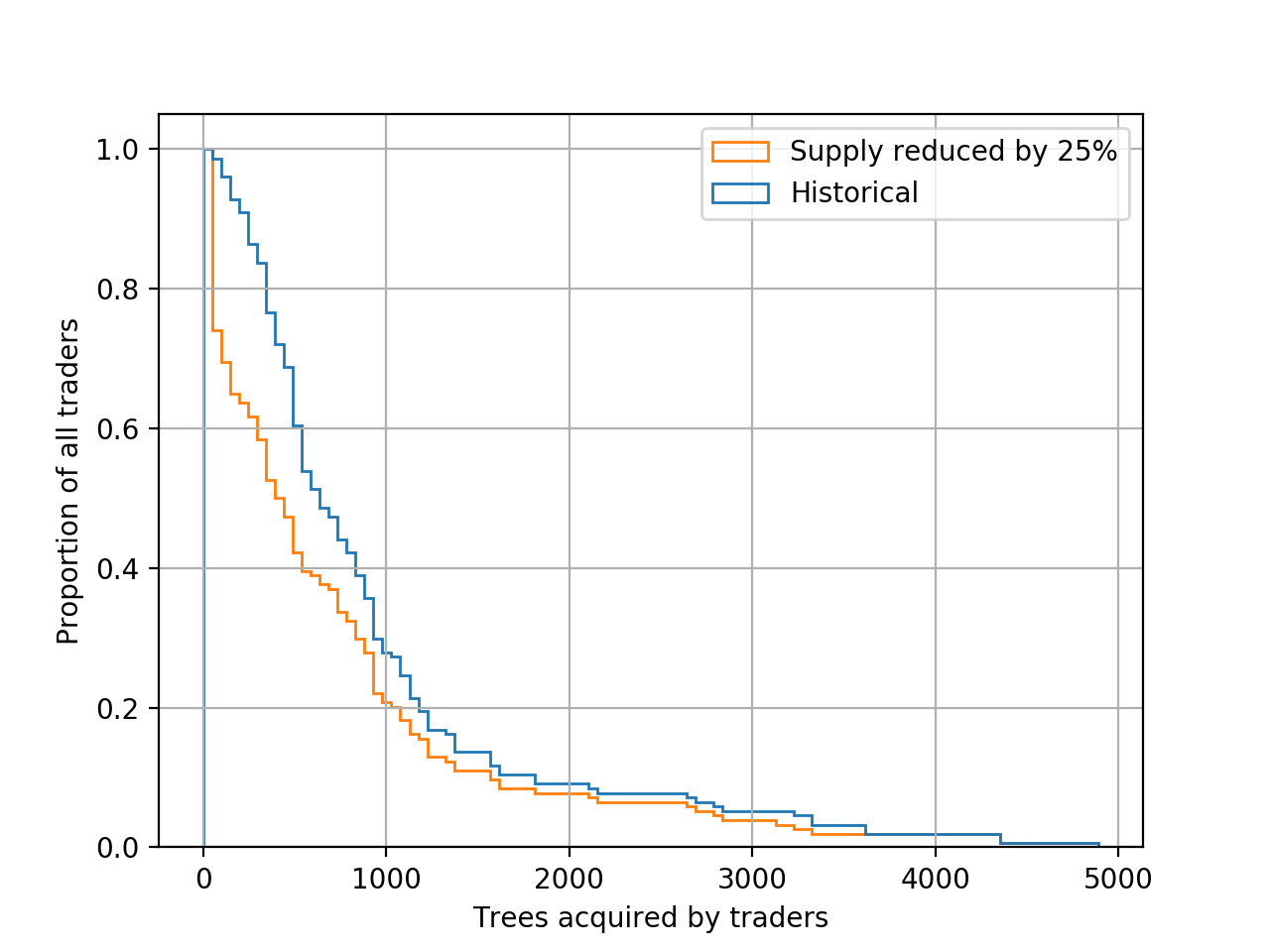}
\caption{Survival function plot of the distribution of the number of trees collected by individual traders based on the actual data (historical) and when the supply is reduced by 25\%.}
\label{fig:new_distribution}
\end{figure}

Regarding the distribution of trees handled by traders, we do not observe any change due to our assumption in the problem setup. We assume the number of trees collected by each trader truly reflects the market demand and set the demand at trader nodes to be the sum of transaction volume going to the trader node according to the data set. The distribution of the number of trees handled by traders is shown by the "Historical" trend line in Fig.~\ref{fig:new_distribution}. As we can see, there are a few traders collecting more than 2,000 trees. This number is considerably higher than the average number of trees collected by other traders and might indicate high market power of a trader. The distribution of the demand would not change as long as the total supply is greater than or equal to total demand. However, when the supply is not sufficient to meet the demand, the distribution is changed significantly. We lower the supply of each village by 25\% and show the distribution for this case compared to the original, with this case shown by the "Supply reduced by 25\%" trend line in Fig.~\ref{fig:new_distribution}. We observe that many traders have no trees collected. One remedy for the model is to increase the lower bound of trees collected by each trader to a positive value. The network with a non-zero lower bound on the flow can be transformed to one with a zero lower bound, and we can then optimize the network flow using our method described above. 

\section{Conclusion}
Our model offers a tool to aid policy- and decision-makers in planning and managing trees on farms and timber markets to promote sustainable production and support farmer livelihoods. In comparing the optimization model with actual data, we demonstrate the potential to reduce costs associated with harvesting and transferring the timber within this region. By reducing the costs, we realize an opportunity to support tree planting in the region to restore over-harvested farms and maintain a level of tree supply to meet market demand. We show which villages to target for tree planting programs to establish an optimal flow regime. With the optimality strategy, we show the potential to eliminate destructive tree harvest, manage trees effectively, and support farmer’s tree-based livelihoods by creating a more equitable and efficient market. Additional or different optimality conditions may be applied to our model to vary the constraints or the outcome priorities based on policy-maker goals. 

Finally, we note the apparent price of anarchy in the timber market of Bilaspur District \cite{PhysRevLett.101.128701}. Here it seems the transaction network optimizes on personal utility for the traders (Nash equilibrium), rather than on the social optimum of our model. In future work, we intend to investigate the sensitivity of breaking or adding individual farmer-trader ties in this network to suggest more refined actions and restrictions that the state forest department may take to help push the Nash equilibrium closer to the social optimum.

\section{Acknowledgements}
This work was funded in part by McIntire-Stennis project no. 1-600418-875000-875912, and in part by the IBM-Illinois Center for Cognitive Computing Systems Research (C3SR), a research collaboration as part of the IBM AI Horizons Network.

\bibliographystyle{IEEEtran}
\bibliography{main}

\end{document}